\documentclass[final,5p,times,twocolumn]{elsarticle}
\usepackage[colorlinks=true,pdfstartview=FitV,linkcolor=blue,citecolor=blue,urlcolor=blue,bookmarks=false]{hyperref}
\usepackage[usenames,dvipsnames]{color}
\usepackage[sc]{mathpazo}
\usepackage[below]{placeins}
\usepackage{afterpage}
\usepackage[separate-uncertainty,retain-explicit-plus,per-mode = symbol, detect-weight=true, range-phrase=-, range-units=single]{siunitx}
\usepackage{longtable}
\usepackage{multirow}
\usepackage{rotating}
\usepackage[mathlines]{lineno}
\usepackage{graphicx}
\usepackage{paralist}
\usepackage{wrapfig}
\usepackage{appendix}
\usepackage{vruler}
\usepackage{etoolbox}
\usepackage{tikz}
\usepackage{listings}
\usepackage{amsmath}
\usepackage{subfigure}
\usepackage{relsize}
\usepackage{etoolbox}
\usepackage{gensymb}
\usepackage{colortbl}
\modulolinenumbers[5]
\newcolumntype{d}[1]{D{.}{\cdot}{#1}}
\newcommand{\ur}{$^{238}$U}
\renewcommand\th{$^{232}$Th}
\newcommand{\knat}{$^{nat}$K}

\newcommand{\vov}{v/v}

\newcommand{\dupont}{DuPont}

\DeclareSIUnit\bqkg{Bq/kg_\text{Ar}}
\DeclareSIUnit\ppt{pg/g}
\DeclareSIUnit\ppb{ng/g}
\DeclareSIUnit\ppm{\ensuremath{\micro}g/g}
\DeclareSIUnit\gpg{g/g}
\DeclareSIUnit\c{$c$}
\DeclareSIUnit\day{day}
\DeclareSIUnit\week{w}
\DeclareSIUnit\year{yr}
\DeclareSIUnit\standard{std}
\DeclareSIUnit\str{sr}

\pdfoutput=1
\newif\ifcolorfigs
\colorfigstrue        

\makeatletter
\def\ps@pprintTitle{%
 \let\@oddhead\@empty
 \let\@evenhead\@empty
 \def\@oddfoot{}%
 \let\@evenfoot\@oddfoot}
\makeatother

\journal{}

\biboptions{sort&compress}

\begin{document}
\begin{frontmatter}
\title{Ultra-low radioactivity flexible printed cables}

\author{Isaac J. Arnquist\corref{cor2}}
\author{Maria Laura di Vacri\corref{cor2}}
\author{Nicole Rocco\corref{cor2}}
\author{Richard Saldanha\corref{cor2}}
\author{Tyler Schlieder\corref{cor2}}
\address{Pacific Northwest National Laboratory, Richland, Washington, 99352 USA}
\author{Raj Patel, Jay Patil, Mario Perez, Harshad Uka}
\address{Q-Flex Inc., Santa Ana, California, 92705 USA}
\begin{abstract}
Flexible printed cables and circuitry based on copper-polyimide materials are widely used in experiments looking for rare events due to their unique electrical and mechanical characteristics. However, past studies have found copper-polyimide flexible cables to contain 400-4700 pg $^{238}$U/g, 16-3700 pg $^{232}$Th/g, and 170-2100 ng $^{nat}$K/g, which can be a significant source of radioactive background for many current and next-generation ultralow background detectors. This study presents a comprehensive investigation into the fabrication process of copper-polyimide flexible cables and the development of custom low radioactivity cables for use in rare-event physics applications.  A methodical step-by-step approach was developed and informed by ultrasensitive assay to determine the radiopurity in the starting materials and identify the contaminating production steps in the cable fabrication process. Radiopure material alternatives were identified, and cleaner production processes and treatments were developed to significantly reduce the imparted contamination. Through the newly developed radiopure fabrication process, fully-functioning cables were produced with radiocontaminant concentrations of 20-31 pg $^{238}$U/g, 12-13 pg $^{232}$Th/g, and 40-550 ng $^{nat}$K/g, which is significantly cleaner than cables from previous work and sufficiently radiopure for current and next-generation detectors. This approach, employing witness samples to investigate each step of the fabrication process, can hopefully serve as a template for investigating radiocontaminants in other material production processes.
\end{abstract}
\end{frontmatter}
\section {Introduction} 
\label{sec:intro}
Cabling associated with signal sensors and readout electronics are often a significant contributor to the radioactive background budget of physics experiments looking for rare events, such as searches for neutrinoless double beta decay or the direct detection of dark matter \cite{adhikari2021nexo, aguilar2022oscura, leonard2017trace, aguilar2022characterization, adams2021sensitivity, agnese2017projected, armengaud2017performance, abgrall2021legend}. Circuitry and cables are composed of two major components - the conductor (typically copper) and insulator. Polyimides are widely used as an insulating substrate in the electronics industry due to their unique properties of high resistivity, high dielectric strength, and flexibility. Polyimides are also stable across a wide range of temperatures, have good thermal conductivity, a thermal expansion coefficient that is close to copper, and a low outgassing rate, which make them a favorable material for use in the ultra-high vacuum and cryogenic environments that are commonly found in low background experiments. 

However, commercial flexible printed cables composed of laminates of copper and polyimide layers are not very radiopure, with typical contamination levels of the primordial radioactive nuclides $^{238}$U and $^{232}$Th measured at a few thousand and a few hundred parts-per-trillion by mass (ppt or \si{\pico\gram\per\gram}), respectively. These values often exceed the stringent radioactivity requirements of rare event searches and experiments therefore have to either limit the amount of cabling used to the absolute minimum necessary (as in nEXO \cite{kharusi2018nexo}, OSCURA \cite{aguilar2022oscura}), or use other materials \cite{busch2018low, andreotti2009low} that, while more radiopure, do not have all the advantageous properties of polyimide-based flexible cables, such as ease of clean and reliable installation. Additionally, since polyimide-based cables are an industry standard for electronics, the use of alternative material often requires custom-made components, which can increase cost, risk, and production time.

Efforts have previously been made to reduce the contamination in copper-polyimide cables used for rare event searches. The EXO-200 collaboration invested significant effort in selecting amongst different commercial copper-polyimide laminates, measuring contamination levels during the photolithography process, investigating alternative chemicals, as well as working with vendors to reduce contamination during handling and packaging \cite{pocar_slides, auger2012exo}. The EDELWEISS collaboration also surveyed different commercial laminates and adhesives to identify low radioactivity materials and used a custom fabrication process \cite{zhang2015novel, armengaud2017performance}. Even with these efforts, the final cables used in the experiment contained \ur~and \th~contamination at the level of hundreds to thousands of ppt. 

Some of the authors of this paper previously worked to identify clean copper-polyimide laminates, in collaboration with \dupont, leading to the production of custom material that had \ur~contamination levels $< 10$ ppt, roughly $10\times$ cleaner than commercial options \cite{arnquist2020ultra}. However, as described in this paper, even when starting with radiopure laminates the fabrication of the cables introduces large amounts of contamination, leading to roughly the same levels found in commercial cables, i.e., a few thousand ppt \ur~and a few hundred ppt \th.

In this paper we systematically investigate sources of contamination in the existing cable production process and study alternative strategies --- such as modifying processes, developing new cleaning methods, and changing sources of raw materials --- to minimize radioactivity in the end product cable.
\section{Cable Fabrication Process and Research Approach}
\label{sec:photolith}
\begin{figure*}[t]
    \centering
    \includegraphics[width=\textwidth]{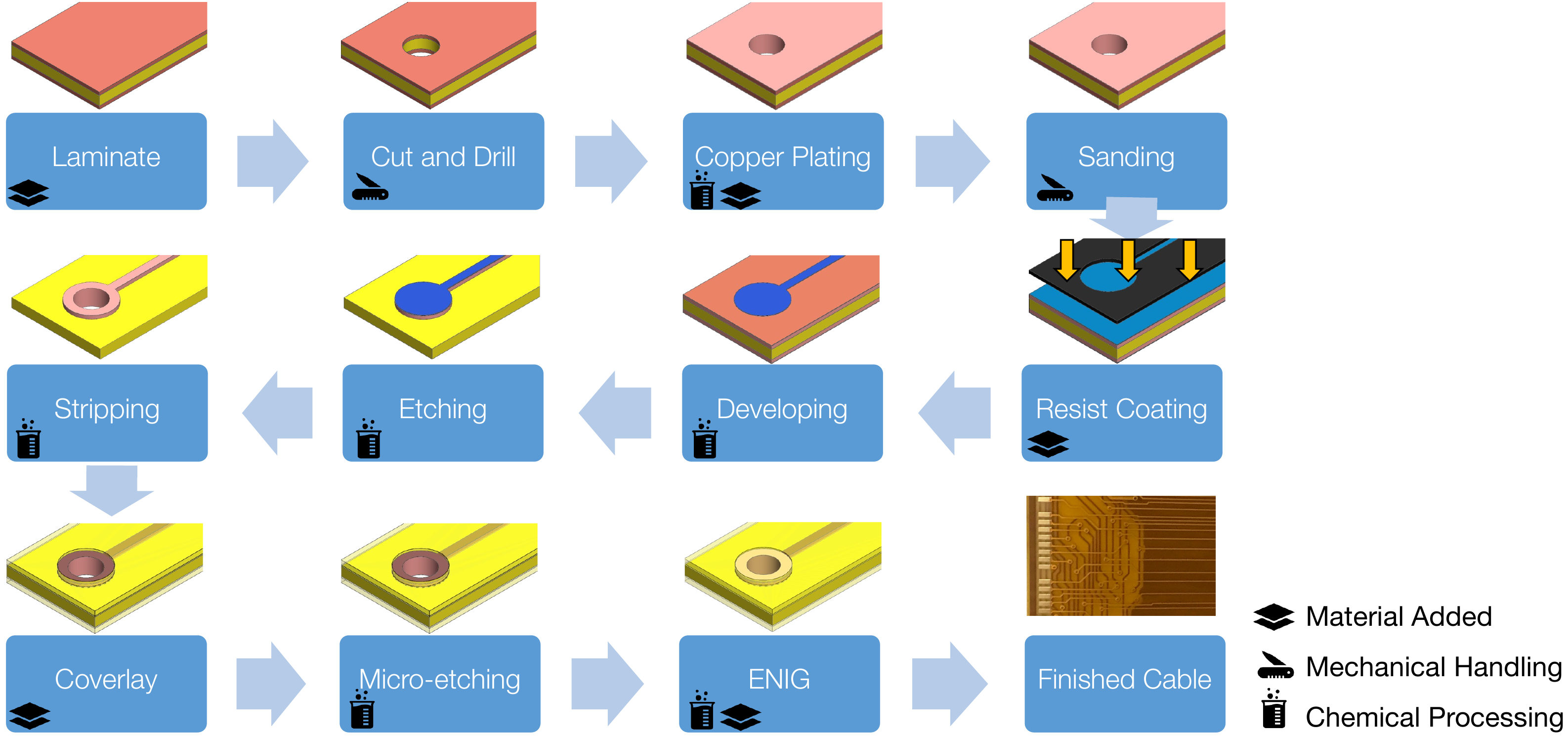}
    \caption{Schematic showing the typical fabrication steps for a two-layer flexible cable. The figures associated with each step show an example of what a simplified cable with a single trace and via looks like after the step is completed. Icons (with legend on the far right) indicate possible vectors of contamination at each step.}
    \label{fig:fab_process}
\end{figure*}

All cable production work was spearheaded by Q-Flex Inc. who specialize in the fabrication of custom long flexible cables. The production of flexible cables involves a number of fabrication steps that start from a simple laminate of copper and polyimide and then adds vias, patterns of traces and interconnections, insulating coverlays, and metallized pads for external connections. The typical commercial photolitohographic process for producing a two-layer flexible cable is shown in Figure~\ref{fig:fab_process}. As part of the fabrication process, the starting laminate is subjected to various chemicals, addition of additional copper and polyimide layers, and mechanical manipulation (drilling, cutting, lamination, etc.). Each one of these steps is a potential source of radioactive contamination. 

\begin{figure}
\centering
 \includegraphics[height=0.8\columnwidth, angle=90]{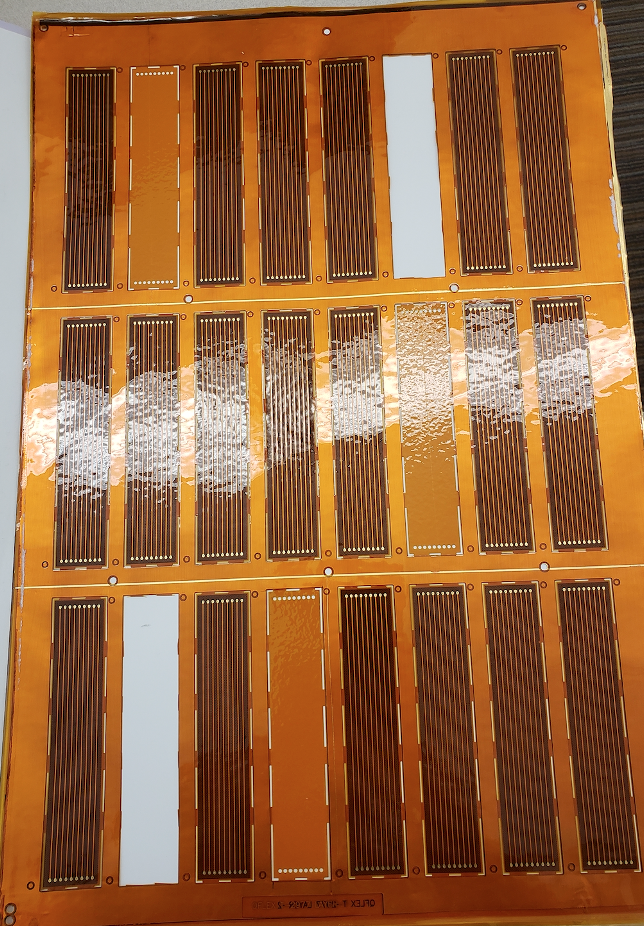}
\caption{An example of a panel of detachable coupons with typical cable features, utilized to study the contamination levels at each step of the fabrication process.}
\label{fig:panel}
\end{figure}

To investigate the contamination levels at each step of the fabrication process, we produced panels of small detachable “coupons” (Figure \ref{fig:panel}). The coupons act as surrogates for real cables – they contain all the relevant features such as traces, vias, coverlays, and go through the same production steps as regular cables. The advantage of this approach is that individual coupons can be easily detached after each step in the process and measured for radioactive contaminants. By comparing the contamination levels in each successive coupon one can assess the level of contamination imparted from each step towards production of the final cable. Where a spray or chemical bath was used for a particular production step, aliquots of the chemicals were also collected for assay along with the coupon for that step. 

Inductively coupled plasma mass spectrometry (ICP-MS) was chosen as the analysis method for measuring the radioactive contamination in each coupon. ICP-MS allows for ultrasensitive assays of materials, processes, and reagents, and the ICP-MS capability at Pacific Northwest National Laboratory (PNNL) for supporting ultralow background physics has developed a variety of methods to reach sub-ppt sensitivities for \ur~,  \th, and $^{nat}$K in a variety of materials \cite{arnquist2017mass,arnquist2020ultra,arnquist2020automated,arnquist2017quick}. ICP-MS also has the distinct advantage of being able to investigate small samples ($<$ 1 gram) with sufficient sensitivity and having quick turnarounds of sample batches, allowing for detailed, systematic investigations of a large number of samples. A detailed description of the ICP-MS capability and method  is given later in this paper. 
Because ICP-MS testing is destructive, the technique cannot be used to measure the actual final cables that are deployed. By adding these small detachable surrogate coupons to the same panel as the cable, we can sample the contamination during the production of the final cables – enabling verification that the cables meet radiopurity requirements. 

ICP-MS directly detects the atoms (ions) within a sample and is therefore most sensitive to those radioisotopes with long half-lives (e.g., \ur~ and \th). In order to measure contamination levels of isotopes further down in the chain, one could use gamma counting of the radioactive decays. However, the sensitivity is typically limited to specific activities at the level of \si{\milli\becquerel\per\kg} (equivalent to 81 ppt \ur, 246 ppt \th), and it requires kilograms of material and counting times on the order of a month. It should be noted that comparison of ICP-MS measurements of the \ur~levels and gamma counting of $^{226}$Ra for cables used in the DAMIC experiment indicated secular disequilibrium, i.e., the activity of $^{226}$Ra in the lower half of the chain was found to be significantly lower than the \ur~activity at the top of the chain \cite{aguilar2022characterization}. We do intend to eventually measure our final cables through gamma counting but the limited sensitivity, large sample mass and long counting times needed, led to us not pursuing it at the R\&D stage.
\section{Analysis Methods}
\label{sec:analysis}
All analyses of the materials reported in this work were performed at PNNL. Analytical work was performed in a Class 10,000 cleanroom, and a laminar flow hood providing a Class 10 environment was used for sample preparations. Details on the chemical reagents used and the preparation of all labware prior to sample handling are given in \ref{sec:labware}.

\subsection{Sample digestion}\label{sec:digestion}

Subsamples of $\sim$\SI{50}{\milli\gram} were cut from original sample with clean stainless steel scissors, collected, and weighed in validated perfluoroalkoxy alkane (PFA) screw cap vials from Savillex (Eden Prairie, MN). Unless explicitly stated, only central portions of the coupons/cables were used for subsampling, excluding any regions with vias, pads, or ENIG processing. Low background clean plastic tongs were used to transfer subsamples to validated polytetrafluoroethylene PTFE iPrep$^{TM}$ vessels for microwave-assisted digestion in a Mars 6 system (CEM Corporation, Matthews, NC). All samples and process blanks were spiked with a known amount ($\sim$ \SIrange{100}{200}{\femto\gram}) of gravimetrically measured $^{229}$Th and $^{233}$U radiotracers (Oak Ridge National Laboratory, Oak Ridge, TN) before digestion. Triplicates of each sample and three process blanks were prepared for each batch of analyses. Digestion was performed in \SI{5}{\milli\liter} of concentrated Optima grade HNO$_3$ at \SI{250}{\celsius}. After complete digestion, sample solutions were transferred to cleaned and validated PFA Savillex vials, and properly diluted for ICP-MS analysis.

\subsection{ICP-MS Analysis} 
Determinations of Th and U were performed using either an Agilent 8800 or 8900 ICP-MS (Agilent Technologies, Santa Clara, CA), each equipped with an integrated autosampler, a microflow PFA nebulizer and a quartz double pass spray chamber. Plasma, ion optics and mass analyzer parameters were tuned based on the instrumental response of a tuning standard solution from Agilent Technologies containing \SI{0.1}{\nano\gram\per\gram} Li, Co, Y, Ce, Tl. In order to maximize the signal-to-noise in the high $m/z$ range for Th and U analysis, the instrumental response for Tl, the element in the standard with $m/z$ and first ionization potential closest to those of Th and U, was used as a reference signal for instrumental parameter optimization. Oxides were monitored and kept below $2\%$ based on the $m/z = 156$ and $m/z = 140$ ratio from Ce (CeO$^{+}$/Ce$^{+}$) in the tuning standard solution. An acquisition method of three replicates and ten sweeps per replicate was used for each reading. Acquisition times for monitoring $m/z$ of interest (\textit{e.g.,} tracers and analytes) were set based on expected signals, in order to maximize instrumental precision by improving counting statistics.

Quantitation of $^{232}$Th and $^{238}$U was performed by isotope dilution methods, using the equation:
\begin{equation}
    \text{Concentration} = \dfrac{A_\text{analyte} \cdot C_\text{tracer}}{A_\text{tracer}} 
\end{equation}
where $A_\text{analyte}$ is the instrument response for the analyte, $A_\text{tracer}$ is the instrument response for the tracer and $C_{tracer}$ is the concentration of the tracer in the sample. Isotope dilution is the most precise and accurate method for quantitation for ICP-MS analysis, allowing the verification of sample preparation efficiency, and accounting for analyte losses and/or matrix effects. Absolute detection limits on the order of a few femtograms were attained for both \ur~and \th. Detection limits normalized to sample mass were on the order of single digit \si{\pico\gram\per\gram} for both Th and U, corresponding to \si{\micro\becquerel\per\kilo\gram}  in terms of radioactivity.

Determinations of natural potassium contamination levels were performed in cool plasma with NH$_3$ reaction mode. Instrumental parameters were optimized based on the instrumental response from a solution containing $\sim$ \SI{1}{\nano\gram\per\gram} $^{nat}$K with natural isotopic composition, diluted in-house from standard solutions. Quantitations of potassium were performed using an external calibration curve, generated using in-house diluted standard solutions with natural isotopic composition. Relatively high potassium backgrounds from the microwave digestion PTFE iPrep$^{TM}$ vessels (see Section~\ref{sec:digestion}) limited detection limits on potassium to the \si{\nano\gram\per\gram} range. Due to the extensive preparation required to minimize these backgrounds, potassium determinations were only performed on a limited number of samples.

All reported central values are the average value of multiple replicates (typically n = 3). Uncertainties for individual replicates were determined from the propagated uncertainties of the instrumental precision. The overall uncertainty on the central value was taken to be the larger of the uncertainty from the instrumental precision and the standard deviation of the replicates. Samples for which the analyte concentration was measured below the detection limit are reported as an upper limit. Values are reported in \si{\pico\gram\per\gram} of sample for $^{238}$U and $^{232}$Th, and in \si{\nano\gram\per\gram} of sample for $^{nat}$K. Corresponding activities in \si{\micro\becquerel\per\kilo\gram} of sample can be obtained based on the specific activity and isotopic abundance of the radionuclides. For reference, 1 \si{\pico\gram~^{238}U\per\gram} of sample corresponds to 12.4 \si{\micro\becquerel\per\kilo\gram} of sample, 1 \si{\pico\gram~^{232}Th\per\gram} corresponds to 4.06 \si{\micro\becquerel\per\kilo\gram}, and 1 \si{\nano\gram~^{nat}K\per\gram} corresponds to 30.5 \si{\micro\becquerel\per\kilo\gram}.

\section{Contaminating Steps}
\label{sec:contamination}

\begin{figure*}
    \centering
    \includegraphics[width=0.8\textwidth]{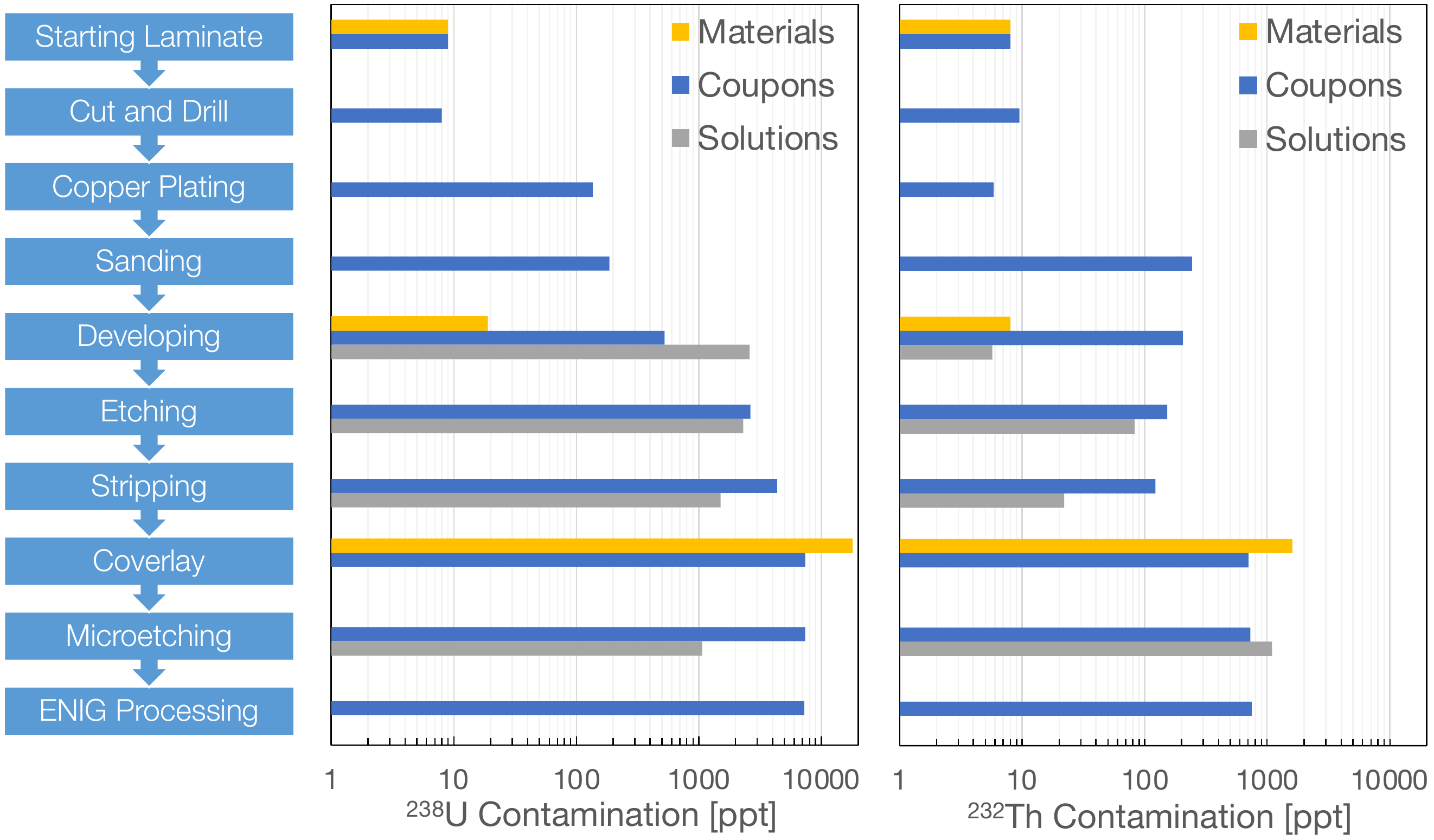}
    \caption{Results of the systematic ICP-MS analysis of coupons, materials, and solutions involved in each step of the fabrication process. The starting laminate used here was Taiflex 2FPDR2005JA, the photoresist was DuPont Riston MM500, and the coverlay was Taiflex FHK1025. For values and uncertainties refer to Table~\ref{tab:init_process_contamination}.}
    \label{fig:init_process_contamination}
\end{figure*}

The measured contamination of \ur~and \th~in coupons at each step of the fabrication process, along with measurements of any materials and solutions involved in those steps, is shown graphically in Figure~\ref{fig:init_process_contamination} with the detailed results and uncertainties presented in Table~\ref{tab:init_process_contamination}. As can be seen, despite starting the process with clean material (few ppt \ur~and \th), the final cable has contamination levels of $\sim$ 7000 ppt \ur~ and $\sim$ 700 ppt \th. These values are similar to the previous findings of the EXO-200 \cite{pocar_slides} and EDELWEISS \cite{zhang2015novel} collaborations. By studying the evolution of the cumulative contamination in the coupons after each step, we were able to target specific production steps that were the dominant source of contamination. In the following subsections we describe each of the key contaminating steps (not necessarily in order) and the approach used to reduce the contamination introduced during that step.

\subsection{Starting Laminate}
The starting material for the production of flexible cables is a laminate, a foil consisting of an insulating flexible polymer, typically polyimide, layered with a conductive metal, usually copper. Since this material forms the bulk of the mass of the finished cable, it is critical that it be as low in radioactivity as possible.

In previous work \cite{arnquist2020ultra}, we have demonstrated the possibility to obtain low-background laminates with single-digit ppt levels of $^{232}$Th and $^{238}$U impurities ($\sim$\si{\micro\becquerel\per\kilo\gram} range). As shown in Table~\ref{tab:commercial_laminates}, this is significantly cleaner than other laminates we surveyed as part of that study as well as laminates measured by other low background experiments. The specialized Kapton laminate produced by the DuPont R\&D division \cite{arnquist2020ultra} was only available in short individual sheets, limiting the ability to make long cables. Measurements conducted prior to the rest of the work described in this paper, found that a copper-polyimide laminate from Taiflex \cite{taiflex} has similar radiopurity to that of the Dupont R\&D material (see Table~\ref{tab:commercial_laminates}). The Taiflex material is two-sided copper laminated to a central polyimide layer with a thermoplastic polyimide. The central polyimide layer is \SI{50}{\micro \meter} and each copper layer is \SI{18}{\micro \meter} thick. This commercial material is available in rolls, making it suitable for manufacturing long cables, as is often needed in low background experiments. All samples made for testing subsequent steps in the process were made starting from this Taiflex laminate.

\begin{table*}[t]
\centering
\begin{tabular}{l c c c c c c}
\hline
\textbf{Type} & \textbf{Vendor} & \textbf{Polyimide} & \textbf{Copper }  & \textbf{\ur} & \textbf{\th} & \textbf{\knat} \\ 
 & & \textbf{Thickness} & \textbf{Thickness} & & & \\
  & & \textbf{[\si{\micro\meter}]} & \textbf{[\si{\micro\meter}]} & \textbf{[\si{\ppt}]} & \textbf{[\si{\ppt}]} & \textbf{[\si{\ppb}]} \\
\hline
Taiflex 2FPDR2005JA  & Taiflex & 50.8 & 18 (x2) & \num{8 \pm 6} & \num{9 \pm 4} & $<$ 100 \\
300ELJ+Cu laminate \cite{arnquist2020ultra} & \dupont~R\&D & 76.2 & 18 (x2) & \num{8.6 \pm 3.6} & \num{20 \pm 14} & \num{164 \pm 82} \\
Pyralux AP8535R \cite{arnquist2020ultra} & \dupont & 76.2 & 18 (x2) & \num{158.0 \pm 6.1} & \num{24.1 \pm 0.9} & $<$ 210\\ 
Novaclad 146319-009 \cite{arnquist2020ultra} & Sheldahl & 50.8 & 5.0 (x1) & \num{283 \pm 21} & \num{50.1 \pm 3.9} & $<$ 210\\
Cirlex \cite{arnquist2020ultra} & Fralock & 228.6 & 34 (x2) & \num{413 \pm 45} & \num{71.4 \pm 2.1} & $<$ 210\\
Espanex MC15-40-00VEG \cite{leonard2008systematic} & Nippon Steel & 40 & 15 (x1) & \num{121 \pm 32} & $< 250$ & \num{880 \pm 120}\\
Espanex MC18-25-00CEM \cite{leonard2008systematic} & Nippon Steel & 25 & 18 (x1) & $< 46$ & $< 260$ & $< 146$\\
Espanex SB \cite{zhang2015novel} &  & 25 & 18 (x2) & $< 284$ & $< 812$ & $< 1777$ \\
\hline
\hline
\end{tabular}
\caption{Comparison of the radiopurity of copper-polyimide laminates. The (xN) factor  in the fourth row indicates whether the copper was on one or both sides of the laminate.}
\label{tab:commercial_laminates}
\end{table*}

\subsection{Photolithography and Cleaning}\label{sec:cleaning}

Photolithography \cite{lithography} is the patterning of electrical traces and other features on the copper surface of the laminate. A photo-sensitive material (photoresist), in our case DuPont Riston MM500, is applied to the laminate to mask sections of the copper and then hardened (developed) by exposure to ultraviolet light. The exposed copper is then removed (etched) and finally the resist is removed (stripped) and dried. This process involves immersing the coupons in a sequence of different chemical solutions to perform these steps. As can be seen in Figure~\ref{fig:init_process_contamination}, the photolithography steps (``Developing'', ``Etching'', and ``Stripping'') were found to be a major source of contamination with 1000's of ppt \ur~ and 100's of ppt \th~measured after the stripping step. We assayed the chemical solutions used in the developing, etching and stripping steps: anhydrous  Na$_2$CO$_3$ diluted to 1$\%$, 0.5 M HCl, and 
a KOH based solution diluted to 20$\%$, respectively, and found significant levels of contamination (shown in Figure~\ref{fig:init_process_contamination}) indicating that the solutions were the likely source of contamination. The chemicals used in this process have been carefully chosen and balanced for cable production and changing the chemicals (e.g. purifying or replacing with alternatives) was determined to require significant amount of R\&D and to likely be impractically expensive. Instead, we pursued the development of a post-cleaning method to try and reduce the contamination after lithography.

The lithogragraphic process for flexible cable manufacturing is almost entirely a "subtractive" procedure: the electrical traces and connections are produced through removal of copper from the outer layers of the starting copper-polyimide laminate with no additional material added (the photoresist is removed at the end) - see Figure~\ref{fig:fab_process}. With this in mind, it was believed that contaminant species from the processing should be at or near the surface and be removable through well-tailored post-processing cleaning. 

Several different cleaning methods were tested. A partially manufactured test cable, processed up to and including the stripping step in Figure~\ref{fig:init_process_contamination} but skipping the "Copper plating" step, was used for this dedicated study. This test cable had a contamination level of $\sim$ 4000 ppt \ur~ and $\sim$ 100 ppt \th. The cleaning consisted of submerging subsamples of the test coupon in the chosen cleaning solution for a fixed length of time while agitating, followed by rinsing in deionized water and air drying in a class 10 laminar flowhood.  Subsamples of the same size as those used for ICP-MS assay ($\sim$~50 mg, $\sim$~1 cm$^2$ surface) were cleaned and assayed before and after cleaning to assess the efficacy of the method. 

Initial attempts tested combinations of various diluted acid solutions (e.g., nitric acid, sulfuric acid) and durations (30 sec - 15 min) to determine the optimal procedure. Although some of these cleaning methods provided a very significant reduction of contaminants in the cable ($\sim$~75x and $\sim$~2x in \ur~and ~\th~respectively) the residual contamination was still above our target radiopurity levels.

Surveying the literature, we learnt that treatments of polyimide films with alkaline solutions, e.g., ethylenediamine,  resulted in modifications to the chemical structure of the polyimide surface, increasing adhesion \cite{park2012surface}. Based on this finding, a cleaning procedure involving the use of an alkaline solution was developed, targeting a more efficient removal of surface contaminants, in particular from the polyimide surface. The cleaning consisted of submerging the samples in 5$\%$ v/v tetramethylammonium hydroxide (TMAH) for 15 min, rinsing in deionized water (DIW), submerging the samples in a 2$\%$ nitric acid solution for $\sim$~1 min, followed by rinsing in DIW and air drying in a class 10 laminar flow hood. A nylon brush with soft polyester bristles was used for scrubbing the samples while they were submerged in each cleaning solution. The duration of the submersion in nitric acid was set based on previous test results, while the duration of submersion in TMAH was set based on a study of cleaning efficacy and visual inspection of the cables after treatment. The efficiency of the cleaning improved with longer duration, but long exposures to TMAH eventually resulted in visible modification of the copper surface. It was found that repeating the entire cleaning cycle thrice was more effective at reducing the contamination than a single cycle. This cleaning recipe reduced the $^{238}$U and $^{232}$Th contamination in the test coupon to $\sim$20 and $\sim$ 10 ppt, respectively. 

Optical inspection of the cable (at 100, 200, and 500x magnification) after cleaning did not show any visible structural changes of the copper traces. For the cleaning of the final cables, described later in Section~\ref{sec:final_cables}, electrical tests were performed after cleaning to ensure the cable still functioned electrically.

\subsection{Copper plating}
\begin{figure*}
    \centering
    \includegraphics[width=\textwidth]{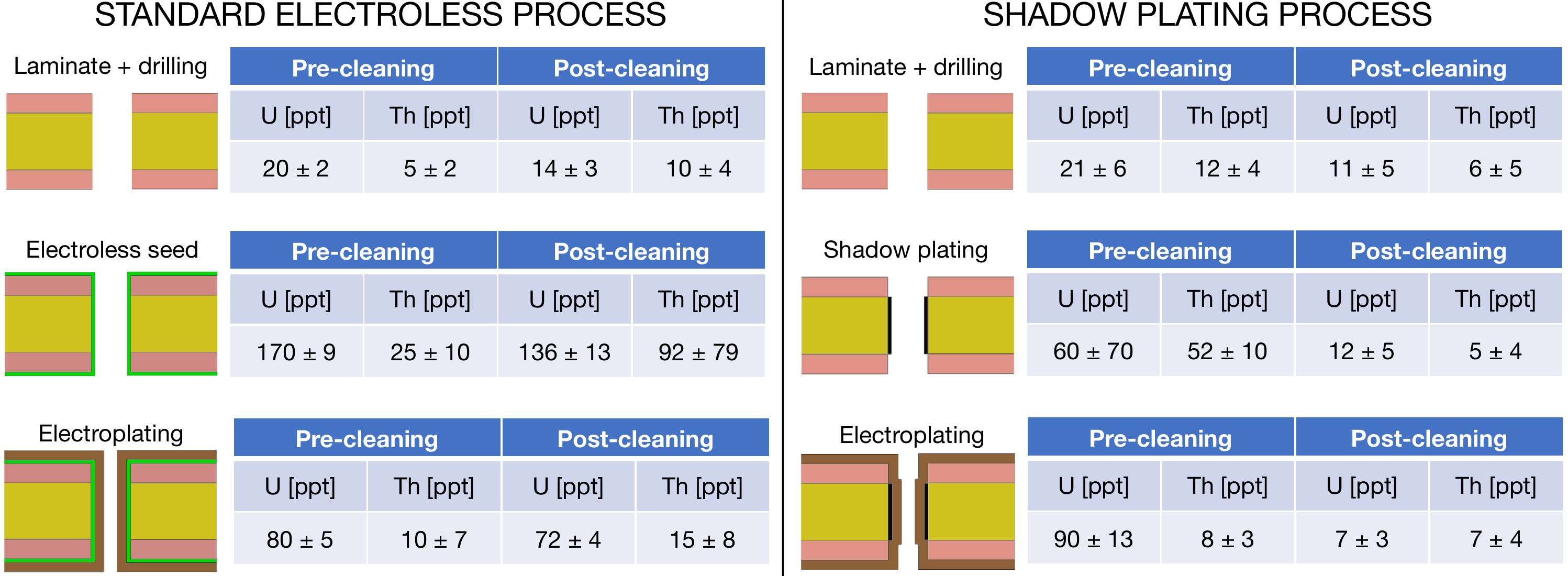}
    \caption{Comparison of standard electroless copper plating process (left) with the shadow plating process (right). In both processes the original laminate of copper (pink) and polyimide (yellow) has the vias drilled out. Left:  A thin copper seed layer (green) is chemically attached to the entire laminate surface. Right: A thin carbon layer is attached to only the polyimide layer, making it conductive. After these steps, in both processes a thicker copper layer (brown) is electroplated on to all conductive surfaces. The tables alongside each step show the measured \ur~and \th~contamination at that step, before and after the cleaning is applied.}
    \label{fig:electroplating}
\end{figure*}
For cables that require interconnections between the conductive copper layers, vias are used. A hole is drilled and then plated with copper to form the electrical connection between the outer two copper layers, across the polyimide insulation (Copper plating step in Figure \ref{fig:init_process_contamination}). Because copper cannot be directly electroplated onto polyimide, a first “seeding” step is used to chemically attach a thin conductive layer to the entire laminate surface. Once the conductive seed layer is added, copper is deposited through electroplating (see the left side of Figure~\ref{fig:electroplating} for a schematic overview). 

Despite several attempts, it was found that the surface cleaning methods we developed (described in Section~\ref{sec:cleaning}) were not effective in removing the contamination introduced during electroplating - see the tables on the left side of Figure~\ref{fig:electroplating} for the results of the measurements before and after cleaning. The electroless seeding process is not selective and deposits the conductive layer over the entire surface of the cable. The subsequent copper electroplating then also deposits a layer of copper over the entire cable. Thus contamination introduced in these steps can be trapped between the base laminate, seed, and electroplating layers, preventing removal through surface cleaning. 
 
To investigate whether the contamination was being trapped between the seed and electroplated layers, we made coupons using a ``pad plating" process \cite{allflex_plating}. In this method, after the seed layer is applied, photoresist is applied to mask the areas of the cable that do not need an additional copper layer (i.e. everything except the vias). The electroplating then only deposits copper in the via regions and the photoresist is removed after the plating. This is a standard industrial process used to avoid the additional plated copper from covering the entire surface area and stiffening the cable. A systematic analysis of coupons after each step of the process, found similar contamination levels with and without pad plating, indicating that the dominant contamination was not trapped between the seed and electroplated layers, but likely between the base laminate and seed layer. 
 
The standard electroless seeding process involves several chemical treatment steps and typically requires a catalyst \cite{ghosh2019electroless}. A newer alternative process to the electroless Cu step is the “shadow” process \cite{allflex_vias, macdermid_shadow}. In this approach, a thin layer of carbon is laid down on the cable. An etch is then performed so that carbon only adheres to the polyimide and so does not cover the copper surface \cite{macdermid_carbon} (see the right side of Figure~\ref{fig:electroplating} for a schematic overview). Unlike the standard electroless seed process, the shadow process uses fewer chemicals and also deposits less material over much smaller area than the standard approach. The shadow process therefore has the potential to be intrinsically less dirty than the standard electroless seed method. 

We investigated the contamination levels in coupons before and after the shadow process as well as after the subsequent electroplating (see tables on right side of Figure~\ref{fig:electroplating}). We found significant contamination after both seeding and plating steps, however, unlike in the case of the electroless seeding, the contamination could be removed by our cleaning method. In fact, the resulting coupons had contamination levels consistent with the base laminate level, roughly 9 times cleaner in \ur~than the electroless seed process. Given these promising results, we elected to use the shadow plating process for copper plating vias in our final cables.

\subsection{Sanding}
 Prior to the application of photoresist, the copper surface of the cable is prepared for optimal dry film adhesion and subsequent clean release. Surface preparation processes typically aim to remove any surface imperfections after drilling and electroplating and roughen the surface to increase film contact area. At Q-Flex, this is done by mechanical scrubbing of the surface with abrasive pads. The scrubbing process was found to increase $^{232}$Th contamination, presumably due to the implantation of small amounts of the abrasive material into the laminate surface. With this in mind, we worked to remove the abrasive material from the surface by sonication in 18.2 M$\Omega\cdot$cm water. The results before and after sonication were within error of each other, indicating that this method did not remove the contamination. The acid-base cleaning method described in Section~\ref{sec:cleaning} was also applied at this step but did not significantly reduce the contamination.
 
 Since our attempts to remove any embedded material from the surface through sonication and cleaning proved ineffective, we reviewed the available commercial options for scouring pads. We decided to exclusively use commercial Scotch Brite$^{TM}$ pads made from SiC \cite{scotch_brite_sic}, known to generally be a radiopure material, rather than previously used pads that used aluminum oxide, titanium dioxide, and other fillers and pigments \cite{scotch_brite_al}. The change in abrasive material led to roughly a 10x reduction in $^{232}$Th at this step.

\subsection{Coverlay}
\begin{table*}[t]
    \centering
    \begin{tabular}{lcccccc}
        \hline
         Sample & PI Thick. & Adh. Thick. & Notes & \ur & \th & \knat\\
         & \textbf{[mil]} & \textbf{[mil]} & & \textbf{[\si{\ppt}]} & \textbf{[\si{\ppt}]} & \textbf{[\si{\ppb}]}\\
         \hline
         Taiflex FHK1025 & 1 & 1 & \multirow{3}{*}{Use epoxy adhesive} & \num{18000 \pm 2000} & \num{1600 \pm 140} & \\
         ShinEtsu CA 333 \cite{leonard2017trace} & 1 & 1 &  & \num{5179 \pm 424} & $< 242$  & \\
         ShinEtsu CA 335 \cite{leonard2017trace} & 1 & 1.4 &  & \num{12020 \pm 390} & \num{9370 \pm 340 }  & \\
         \hline
         Dupont LF0110 & 1 & 1 & \multirow{5}{*}{Use acrylic adhesive} & \num{314 \pm 13} & \num{49 \pm 8} & \num{4000 \pm 2000}\\
         Upilex C120 & 2 & 1 &  & \num{30 \pm 2} & \num{280 \pm 20} & \num{21300 \pm 300} \\
         Panasonic MCL Plus 110 & 1 & 1 & & \num{78 \pm 4} & \num{45 \pm 7} & \num{5030 \pm 140}\\
         Dupont FR 70001 \cite{leonard2017trace} & 0.5 & 0.5 &  & $< 1065$ & $< 473$  & \\
         Dupont FR 0110 \cite{leonard2017trace} & 1 & 1 &  & $< 818$ & $< 273$  & \\
         \hline
         Dupont LF0100 & 0 & 1 & Adhesive in LF0110 & \num{16 \pm 4} & \num{39 \pm 11} & \\
         Imitex MI-100 & 0 & 1 & Adhesive & \num{9 \pm 5} & $<$ \num{14} & \\
         \hline
         \hline
    \end{tabular}
    \caption{Radiopurity results from survey of commercially available coverlay materials, including results from previous measurements in the literature \cite{leonard2017trace}. The second column indicates the polyimide (PI) thickness and the third column the adhesive thickness. The materials in the last two rows are adhesives only.}
    \label{tab:coverlays}
\end{table*}

A coverlay is an insulating layer that is applied over the outer surfaces of a cable to prevent oxidation and shorting of the exposed copper traces. Commercial coverlays are typically composed of two materials — the polyimide that produces the electrical insulation and an adhesive that bonds the polyimide to the surface of the cable. In addition to reviewing previous measurements of coverlays in the literature \cite{leonard2017trace}, we surveyed several different commercially available coverlays, as shown in Table~\ref{tab:coverlays}. 

As can be seen from the results, the contamination levels of \ur~and \th~can vary by several orders of magnitude amongst commercial coverlays. Although Taiflex copper-polyimide laminates were found to be extremely radiopure, the epoxy-based adhesives they utilize have extremely high levels of radiocontaminants. Our measurements indicate that coverlays based on acrylic adhesives have the lowest \ur~and \th~contamination. We also measured the \knat~contamination levels in these acrylic coverlays and found high contamination levels ($>$ ppm) with significant variations (see final column of Table~\ref{tab:coverlays}). Overall, the Panasonic MCL coverlay was the best choice, with contaminations of \ur~and \th~lower by factors of roughly 200 and 30, respectively, compared to the Taiflex coverlay. Upilex C120 should also be considered for experiments that are significantly more sensitive to backgrounds from \ur~than \th~or~$^{40}$K.  
We also separately measured acrylic adhesives and found them to be cleaner than the combined acrylic + polyimide coverlays (last two rows of Table~\ref{tab:coverlays}). For future work we intend to investigate coverlays made from custom lamination of radiopure acrylic and polyimide as well as all-acrylic coverlays.
\section{Production of an ultra-low background cable}
\label{sec:final_cables}
\begin{figure*}
    \centering
    \includegraphics[width=\textwidth]{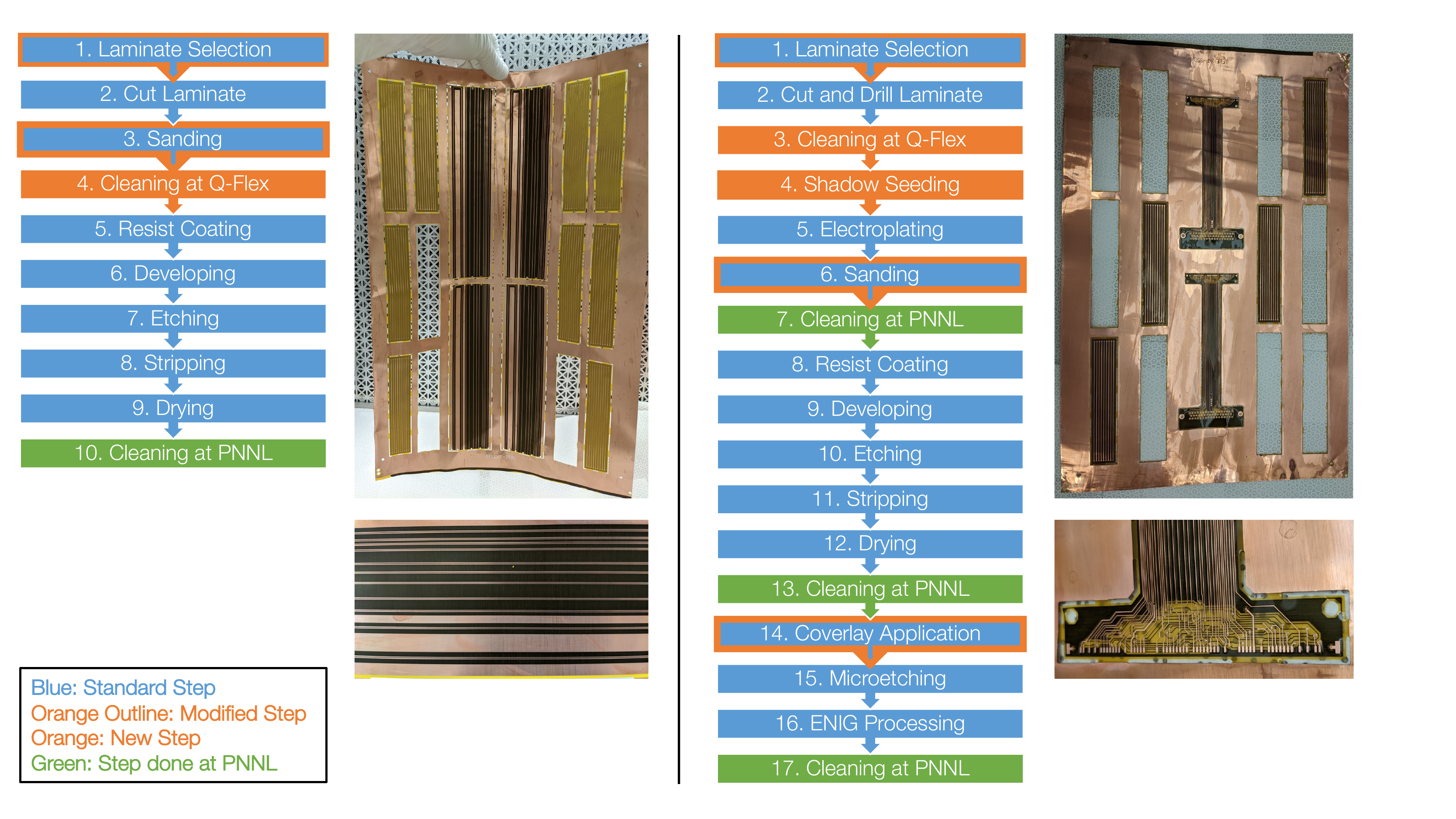}
    \caption{Outline of custom fabrication process and pictures of the panels used for the final SiPM (left) and CCD (right) cables. The color code for the fabrication process flowchart is described in the legend at the bottom left. The top picture shows an entire sample panel with the cables in the center and detachable coupons on the edges. The bottom picture shows a close-up view of the traces and other features on the cables.}
    \label{fig:final_cables}
\end{figure*}

Having identified the key contaminating steps in flexible cable fabrication and developed methods and alternatives to reduce the contamination in each individual step, we  produced two sets of fully-functional flexible cables by combining all the modifications described in the previous section into a single low-background fabrication process. We chose two different flexible cable designs that are being considered for on-going and future nuclear and particle physics experiments. The first design is under development for the readout of silicon photomultipliers (SiPMs) in the nEXO neutrinoless double beta decay experiment \cite{kharusi2018nexo}. Based on findings from previous investigations into radiopure cable conducted for the EXO-200 cable \cite{pocar_slides,leonard2017trace}, the cable design was simplified to keep contamination levels as low as possible. The cable (shown on the left in Figure~\ref{fig:final_cables}) consists of traces on only one side of the cable and a continuous ground plane on the other. As all the traces are on one side of the cable with no interconnections between the two layers, no vias or copper plating is required. Additionally, while it would be advantageous, the design does not include a coverlay or surface metallization of contacts in order to keep contamination levels to a minimum. 

The second cable design was developed for the readout of charged couple devices (CCDs) for the DAMIC-M \cite{arnquist2023first} dark matter experiment. The CCD cable (shown on the right in Figure~\ref{fig:final_cables}) consists of multiple traces and interconnections between the two copper layers that require copper-plated vias. The cable also includes coverlays on both sides of the cable and ENIG metallization on the pads. 

The SiPM and CCD cable designs were chosen because they represent two ends of the spectrum in terms of complexity of two-layer flexible cables. Figure~\ref{fig:final_cables} shows the process steps used to fabricate these two cables, highlighting the modifications made to the standard process. As before, panels consisting of the cables along with several detachable coupons were used and coupons were removed after several key steps so that the contamination levels could be tracked, if needed. The cleaning process described in Section~\ref{sec:cleaning} was applied at key steps in the fabrication process where it would not interfere with the production and would remove surface contamination before material was permanently added to the cable. Since the cleaning recipe required the use of relatively hazardous chemicals and precise cleaning procedures, at intermediate steps (indicated in green in  Figure~\ref{fig:final_cables}) the partially-processed panels were shipped from Q-Flex to PNNL, cleaned, and then shipped back. The shipping was done in sealed plastic bags and the cleaning was done as promptly as possible to avoid oxidation of the copper layers between steps. Once the fabrication steps were complete, the cable was shipped to PNNL for a final cleaning and subsequent measurement. For the measurements of the contamination levels in the final cables, we sub-sampled the cables themselves (not witness coupons) and used 7-8 sub-samples, rather than the standard 3, to increase our confidence that the average of our measurements were representative of the "true" contamination in the cable overall. 

\begin{table}[t]
\centering
\begin{tabular}{c c c c c c }
\hline
Cable & Rep. & \textbf{\ur} & \textbf{\th} & \textbf{\knat} \\
& & \textbf{[pg/g]} & \textbf{[pg/g]} & \textbf{[ng/g]} \\
\hline
\hline
\multirow{9}{*}{\shortstack{SiPM Cable\\(Custom)}} & 1 & \num{20 \pm 2} & \num{< 9.8} & \num{< 38} \\
 & 2 & \num{21 \pm 2} & \num{< 9.4} & \num{< 37} \\
 & 3 & \num{18 \pm 2} & \num{< 8.6} & \num{< 34} \\
 & 4 & \num{20.9 \pm 1.2} & \num{< 10.4} & \num{47 \pm 6} \\
 & 5 & \num{19 \pm 2} & \num{< 10.3} & \num{32 \pm 8} \\
 & 6 & \num{18.8 \pm 1.2} & \num{< 12.3} & \num{< 20} \\
 & 7 & \num{19.6 \pm 1.5} & \num{< 12.0} & \num{52 \pm 7} \\
 & 8 & \num{19 \pm 3} & \num{< 12.0} & \num{28 \pm 7} \\
 \cline{2-5}
 & Avg.* & \num{20 \pm 2} & \num{< 12.3} & \num{40 \pm 12} \\
\hline
\hline
\multirow{8}{*}{\shortstack{CCD Cable\\(Custom)}} & 1 & \num{32 \pm 2} & \num{12 \pm 3} & \num{559 \pm 13}\\
 & 2 & \num{31 \pm 4} & \num{11 \pm 3} & \num{529 \pm 12}\\
 & 3 & \num{29 \pm 2} & \num{< 8.9} & \num{572 \pm 12} \\
 & 4 & \num{32 \pm 3} & \num{16 \pm 4} & \num{569 \pm 13} \\
 & 5 & \num{31 \pm 2} & \num{< 11.7} & \num{558 \pm 12} \\
 & 6 & \num{30 \pm 2} & \num{< 10.9} & \num{546 \pm 9} \\
 & 7 & \num{30 \pm 2} & \num{< 11.1} & \num{515 \pm 9} \\
\cline{2-5}
 & Avg.* & \num{31 \pm 2} & \num{13 \pm 3} & \num{550 \pm 20} \\
\hline
\hline
\end{tabular}
\caption{Measurements of the \ur, \th~and \knat~concentrations in the final SiPM and CCD custom cables. Each row within the cable type lists the values for a given subsample (identified by the repetition number in the second column) with the uncertainties indicating the instrumental uncertainty. 
*Averages are reported as the average and standard deviation of the central values of all replicates for that isotope (upper limits excluded). Where all replicates are upper limits, the largest upper limit is reported.}
\label{tab:cable_final_results}
\end{table}

The results are shown in Table~\ref{tab:cable_final_results}. The levels of contamination measured in the final SiPM cable were extremely low, with roughly 20 ppt of \ur, \th~levels below our sensitivity of $\sim$ 10 ppt, and \knat~at roughly 36 ppt. The CCD cable, which included additional copper plating of the vias, coverlays on both sides, and ENIG metallization of the contacts, was only slightly higher in \ur~and \th, with roughly 31 and 12 ppt respectively. The \knat~levels were significantly higher, consistent with the intrinsic contamination levels measured in the Panasonic MCL Plus 110 coverlay used and the relative contribution of the coverlay to the overall mass of the cable. It is worth remarking that while commercial cables can sometimes have values that vary significantly from subsample to subsample, all subsample replicates measured in the custom final cable were consistent with each other, indicating that the cleaning procedure (applied at key steps) likely removed any non-uniform surface contamination. 

\subsection{ENIG metallization}
While measurements of the center of the cable (where all samples were taken from for the results in Table~\ref{tab:cable_final_results}) represent the contamination levels in the vast majority of the cable material, the copper contacts at the ends of the cable typically have an additional metallization layer to avoid oxidation and improve solderability \cite{macdermid_surface_finish}. This metallization, typically added through electroless deposition in a chemical bath with the addition of a catalyst, can add additional contamination, and though the total mass is often small the contacts at one end of the cable can be located very close to the sensitive volume. The CCD cables were metallized with electroless nickel immersion gold (ENIG) using \SIrange{3.0}{5.6}{\micro\meter} (\SIrange{118}{220}{microinch}) of nickel and \SIrange{0.076}{0.127}{\micro\meter} (\SIrange{3}{5}{microinch}) of gold. In order to measure the contamination levels of the ENIG layer, subsamples were cut from the ends of the cable to include the metallized pads. Since the size of the ENIG region varied by subsample, for each subsample measurement we estimated the ENIG-covered area and subtracted out the average contamination of the rest of the cable layers (estimated from the measurements at the center). The residual contamination was attributed to the application of the ENIG layer. We find contamination levels per unit area of ENIG at roughly \SI{0.1}{\pico\gram\per\milli\meter\squared} for \ur, \SIrange{0.01}{0.1}{\pico\gram\per\milli\meter\squared} for \th, and \SI{10}{\nano\gram\per\milli\meter\squared} for \knat. 
These results can be used to estimate the backgrounds for specific cable layouts with ENIG metallization layers.

\section{Summary and Discussion}
\definecolor{Gray}{gray}{0.9}
\definecolor{LightCyan}{rgb}{0.88,1,1}
\begin{table*}[]
    \centering
    \begin{tabular}{lccccccc}
    \hline
    Cable & Copper & Polyimide  & Coverlay & Surface & \ur & \th & \knat  \\
     & Layers & Layers &  &  Finish & & &  \\
     & [\si{\micro\meter}] & [\si{\micro\meter}] & & &  [\si{\ppt}] & [\si{\ppt}] & [\si{\ppb}]  \\
    \hline
    \rowcolor{LightCyan}
    nEXO SiPM [This Work] & \num{18} (x2) & \num{50.8} (x1) & No & No & \num{20 \pm 2} & \num{< 12.3} & \num{40 \pm 12}\\
    nEXO SiPM [Comm.] & \num{18} (x2) & \num{50.8} (x1) & No & No & \numrange{1300}{6200} & \numrange{16}{63} & \\
    \hline
    \rowcolor{LightCyan}
    DAMIC-M CCD [This Work] & \num{18} (x2) & \num{50.8} (x1) & x2 & ENIG & \num{31 \pm 2} & \num{13 \pm 3} & \num{550 \pm 20}\\
    DAMIC-M CCD [Comm.] & \num{18} (x2) & \num{50.8} (x1) & x2 & ENIG & \num{2600 \pm 40} & \num{261 \pm 12} & \num{170 \pm 50}\\
    \hline
    EXO-200 \cite{pocar_slides, leonard2017trace} & \num{18} (x1) & \num{25.4} (x1) & No & No & \num{412 \pm 47} & $< 117$ & \\
    EDELWEISS III \cite{armengaud2017performance, zhang2015novel}& \num{18} (x4) & 25/125 (x3/x4) & No & No & \num{650 \pm 490} & \num{3700 \pm 2500} & \num{2100 \pm 840} \\
    DAMIC at SNOLAB \cite{aguilar2022characterization}& \num{18} (x5) & \num{25.4} (x4) & x2 & ENIG & \num{4700 \pm 400} & \num{790 \pm 120} & \num{940 \pm 60} \\
    \hline
    \end{tabular}
    \caption{Comparison of the final radiopurity levels achieved in cables produced as part of this work (highlighted in blue) with other commercial and custom cables used in low background rare event experiments. The commercial (abreviated as Comm.) cables listed near the top of the table are the same design as our custom cables and can be compared directly. The cables listed at the bottom of the table have different construction and fabrication options - see text for details.}
    \label{tab:final_comparisons}
\end{table*}

As a direct point of comparison, we also measured commercial cables with the same design as our SiPM and CCD custom cables. For the SiPM cable we measured a cable that was produced with the same low-background starting laminate but using the original standard Q-Flex fabrication process that did not include any of the modifications described in this work. Cables from different batches showed large variations in the \ur~and \th~contamination levels and so we quote a range of values (see Table~\ref{tab:final_comparisons}). For the CCD cable we obtained a cable from the DAMIC collaboration of the same design, fabricated by a different commercial vendor (PCB Universe \cite{pcbuniverse}) with no special instructions on materials to use or process modifications. It can be seen from comparing the results in Table~\ref{tab:final_comparisons} that, compared to our custom low radioactivity cables, \ur~and\th~ levels in the commercial cables are roughly 100$\times$ and 10$\times$ larger, respectively. Interestingly, the potassium contamination levels in the commercial CCD cable were roughly $2\times$ lower than the levels measured in our custom cable, likely due to the choice of a different coverlay material.

It is also  worthwhile to compare our results to those obtained by other experiments that use flat flexible cables in rare event searches, though the cable designs vary significantly. As noted in the introduction, the EXO-200 collaboration, which searched for the neutrinoless double beta decay of $^{136}$Xe, invested significant effort to reduce the contamination in their detector cables \cite{pocar_slides, auger2012exo}. They used single-layer cables and worked with the vendor to use fresh chemicals and containers and add isopropanol rinses during the production process. After fabrication, all small cables were subjected to a post-production plasma etch followed by a cleaning with acetone and ethanol. To reduce contamination, no metallization of the contacts was applied and no coverlay was used within the detector cryostat. The copper and polyimide layers of the cable were separately assayed for radiocontamination using ICP-MS \cite{leonard2008systematic}. Taking the best obtained values for the copper (Entry 261 in Ref.\cite{leonard2008systematic}) and polyimide (Entry 262 in Ref.\cite{leonard2008systematic}) we have combined them by the mass ratio to calculate the final value shown in Table~\ref{tab:final_comparisons}. The EDELWEISS collaboration, searching for dark matter, surveyed several different materials and selected specific materials for their copper clad polyimide laminates \cite{zhang2015novel}. Though they only used four conducting layers, the cable had additional polyimide layers of varying thickness in between to reduce the capacitance. They also identified the photolithography process as an additional source of contamination and used a custom fabrication procedure at the Oxford Photofabrication Unit \cite{zhang2015novel}. The contamination levels in the final cables used in the EDELWEISS III experiment \cite{armengaud2017performance} are listed in Table~\ref{tab:final_comparisons}. The DAMIC collaboration, searching for dark matter, used a 5-layer cable to read out their CCDs installed at SNOLAB \cite{aguilar2022characterization}. These were commercial cables with no effort made to reduce the contamination. The contamination levels, measured through ICP-MS, are listed in Table~\ref{tab:final_comparisons}. It can be seen that the radiopurity levels achieved in the cables produced as part of this work, compared to the ones used in previous experiments, are at least an order of magnitude better in \ur~and \th, and lower in \knat~ as well.

This work was motivated by the specific need for low radioactivity cables in either planned upgrades of currently running experiments such as DAMIC-M \cite{arnquist2023first} and SuperCDMS \cite{albakry2022strategy} or in future proposed experiments such as nEXO \cite{kharusi2018nexo}, Oscura \cite{aguilar2022oscura}.  We have demonstrated that simple cable designs can be produced with contamination levels at 20 ppt \ur, $< 12$ ppt \th, and 40 ppb \knat, nearly at the levels found in the raw materials. We believe these levels to be sufficiently radiopure for all the experiments listed above. 

We have also shown that cable designs that require additional features such as interconnected layers, coverlays, and metallization can also be made with only a relatively small increase in the \ur~contamination, though there is a significant increase in \knat~due to the coverlay material. Additional investigations into multi-layer cables, higher radiopurity coverlays, and alternatives to ENIG metallization are planned in the near future. It should be noted that low radioactivity cables can increase the sensitivity of these experiments by not only reducing the background budget but also by  allowing for the use of additional sensors and simpler designs of the sensor readout within the same background budget. 

Finally, while we have focused this work on radiopure cables, the same low radioactivity fabrication investigation could be applied to flexible printed circuits and related applications. For example, in future work we will use a similar approach with the goal of developing low radioactivity superconducting cables with possible applications to cryogenic rare-event searches \cite{hollister2017cryogenics, andreotti2009low}, and to develop low radioactivity printed circuit boards, which have been predicted to be the dominant material within dilution refrigerators impacting decoherence from background ionizing radiation \cite{cardani2023disentangling}.

\section{Acknowledgements}
We thank David Moore for initiating contact between Q-Flex and PNNL and providing early cable designs. We are grateful to the nEXO and DAMIC collaborations for sharing their proposed cable designs and for discussions on the cable radioactive background targets as well as electrical and mechanical requirements. We thank Marcel Conde for electrically testing the cable after fabrication, Andrea Pocar for helpful discussions on work that was previously done for EXO-200, and Cesar Arciniega for helping with some of the graphics for the paper.

This work was partially supported by the U.S. Department of Energy Office of Science, Office of Nuclear Physics, under the Early Career Research Program  and the  SBIR Program (Award No. DE-SC0021547). Pacific Northwest National Laboratory is a multi-program national laboratory operated for the U.S. Department of Energy (DOE) by Battelle Memorial Institute under contract number DE-AC05-76RL01830.
\bibliography{references.bib}
\appendix
\section{Reagents and Labware}
\label{sec:labware}
 Electronic grade tetramethylammonium hydroxide (TMAH), 100\% pure ethanol, and Optima grade nitric, hydrochloric, and sulfuric acids (Fisher Scientific, Pittsburg, PA, USA) were used for sample cleaning and preparation. 18.2 M$\Omega\cdot$cm water from a MilliQ system (Merk Millipore GmbH, Burlington, MA) was used for sample rinsing and in the preparation of reagent solutions. Ultralow background perfluoroalkoxy alkane (PFA) screw cap vials from Savillex (Eden Prairie, MN) were used as sample containers, to collect solutions from microwave digestion vessels and as ICP-MS autosampler vials. \\
All labware involved in sample handling and analysis (vials, microwave vessels, tongs, pipette tips) were cleaned with 2\% \vov~Micro-90 detergent\textsuperscript{\textregistered} detergent (Cole-Parmer, Vernon Hills, IL), triply rinsed with MilliQ water, and leached in Optima grade 25\% \vov~HCl and 6M HNO$_3$ solutions. Following leaching, all labware underwent validation to ensure cleanliness. The validation step consisted of pipetting a small volume of 5\% \vov~HNO$_3$ into each container: 1.5 mL in the PFA vials and 5mL in the iPrep$^{TM}$ microwave vessels. Vials were closed, shaken, and kept at $\SI{80}{\celsius}$ for at least 12 hours. Microwave vessels underwent a microwave digestion run at $\SI{220}{\celsius}$. Tongs and pipette tips were soaked into a 5\% \vov~HNO$_3$ leaching solution ($ca.$ 1.5 mL) for 5-10 minutes. The leachate from all labware was then analyzed via ICP-MS. The validation was performed to assure sufficiently low background for Th and U. Only labware for which Th and U signals were at reagent background levels passed validation. Labware failing validation underwent additional cycles of leaching and validation tests until background requirements were met.

\section{Contamination by Fabrication Step}
\begin{table*}[t]
\centering
\begin{tabular}{c l c c c c c c}
\hline
\multicolumn{2}{c}{Fabrication Step} & \multicolumn{2}{c}{Materials} & \multicolumn{2}{c}{Solutions} & \multicolumn{2}{c}{Coupons}\\
& & \textbf{\ur} & \textbf{\th} & \textbf{\ur} & \textbf{\th} & \textbf{\ur} & \textbf{\th} \\
 & & \textbf{[pg/g]} & \textbf{[pg/g]} & \textbf{[pg/g]} & \textbf{[pg/g]} & \textbf{[pg/g]} & \textbf{[pg/g]}\\
\hline
1 & Laminate* & \num{9 \pm 4} & \num{8 \pm 6} & & & &\\
2 & Cut and Drill & & & & & \num{<16} & \num{<20}\\
3 & Copper Plating & & & & & \num{140 \pm 10} & \num{<20}\\
4 & Sanding & & & & & \num{190 \pm 20} & \num{240 \pm 110}\\
5 & Developing & \num{19 \pm 13}$^{\dagger}$ & \num{8 \pm 6}$^{\dagger}$ & \num{2630 \pm 110} & \num{6 \pm 2} & \num{530 \pm 40} & \num{210 \pm 10}\\
6 & Etching & & & \num{2300 \pm 200} & \num{80 \pm 10} & \num{2600 \pm 600} & \num{150 \pm 40}\\
7 & Stripping & & & \num{1510 \pm 70} & \num{22 \pm 2} & \num{4400 \pm 500} & \num{120 \pm 30}\\
8 & Coverlay* & \num{18000 \pm 2000} & \num{1600 \pm 140} & & & \num{7400 \pm 200} & \num{710 \pm 20}\\
9 & Microetching & & & \num{1060 \pm 60} & \num{1090 \pm 60} & \num{7400 \pm 400} & \num{730 \pm 50}\\
10 & ENIG Processing & & & & & \num{7300 \pm 200} & \num{750 \pm 40}\\
\hline 
\end{tabular}
\caption{\ur~and \th~concentrations at each cable fabrication step for raw materials, photolithography solutions, and representative coupons. Measurements of the materials and solutions are normalized to their individual component masses, while the values listed for the coupons are normalized to the mass of the coupon, including all layers present, at that step. *Laminate and coverlay from Taiflex (2FPDR2005JA and FHK1025, respectively). $^{\dagger}$Photoresist from DuPont (Riston MM500)}

\label{tab:init_process_contamination}
\end{table*}

\end{document}